\documentclass{article}

\usepackage{amsmath}        %AMS的数学公式和符号
\usepackage{amssymb}
\usepackage{euscript}       % 含有手写体
\usepackage{color}          % 含有颜色
\usepackage{tensor}         % 用以产生复杂的张量指标
\usepackage[a4paper,left=3cm,right=3cm]{geometry}  %geometry宏包用意调节页面尺寸
\usepackage{amsthm}
\title{Generalized Vaidya spacetime for cubic gravity}
\author{Shan-Ming Ruan\thanks{ruanshanming@itp.ac.cn}\\State Key Laboratory of Theoretical Physics, Institute of Theoretical Physics, \\Chinese Academy of Sciences, Beijing 100190, China }
\date{}

\begin{document}

\bibliographystyle{unsrt}   %定义参考文献的类型，一般哪有palin ， unsrt , alpha ,abbrv

\maketitle           %输出论文标题
\begin{abstract}     %输入的abstract
   We present a kind of generalized Vaidya solutions of a new cubic gravity in five dimensions whose field equations in spherically spacetime are always second order like the Lovelock gravity. We also study the thermodynamics of its apparent horizon and get its entropy expression and generalized Misner-Sharp energy. Finally we present the first law and second law hold in this gravity. Although all the results are analogue to those in Lovelock gravity, we in fact introduce the contribution of new cubic term in five dimensions where cubic Lovelock term is just zero.
\end{abstract}

\section {Introduction}
It is generally believed that General Relativity is just an effective model at low energy despite of its success in large scale and it should be replaced by an unknown theory called Quantum Gravity in the UV regime. Before we find the complete quantum gravity, it is useful for us to study some toy model by introducing some higher curvatures terms to GR, for example the Guass-Bonnet term which naturally appears in low energy effective action of heterotic string theory\cite{Zumino:1985dp}\cite{Gross:1986iv} and also can be derived by compactifying six dimensions in compact Calabi-Yau($CY_6$) threefold from $R^4$ term in the bosonic action of the eleven-dimensional supergravity limit of M theory \cite{Garraffo:2008hu}\cite{Antoniadis:1997eg}. On the other hand, Gauss-Bonnet gravity could also be considered as a special Lovelock gravity \cite{Lovelock:1971yv} that is a natural generalization of general relativity in higher dimensions. The Lagrangian of Lovelock gravity is constituted of some extended Euler density and can be written as:
 \begin{equation}
\mathcal{L}= \sqrt{-g}\sum_{i=0}^p \alpha_i \mathcal{R}^i ,
 \end{equation}
where $n$ is the dimension of spacetime, $ p\leq [(n-1)/2] $, $\alpha_i$ are coupling constants with dimension of $[\text{length}]^{2i-2}$.  $\mathcal{R}^i$ are generally considered as Euler densities :
\begin{equation}
 \mathcal{R}^i=\frac{1}{2^i}\delta^{\mu_1\nu_1\mu_2\nu_2\ldots\mu_i\nu_i}_{\alpha_1\beta_1\alpha_2\beta_2\ldots\alpha_i\beta_i}\prod^i_{r=1}R\indices{^{\alpha_r\beta_r}_{\mu_r\nu_r}}.
\end{equation}
If we ignore the boundary term, we can define 2i-dimensional Euler number as
\begin{equation}
  \chi(M)= \frac{(-)^{i+1}\Gamma(2i+1)}{2^{i+2}\pi^i\Gamma(i+1)}\int_M d^{2i} x \sqrt{-g}\mathcal{R}^i .
\end{equation}
A significant feature of Lovelock gravity is that its equations of motion are only second order with respect to metric, which could be easily found by varying the Lagrangian (1) :
\begin{equation}
  \mathcal{G}_a^b =  \frac{1}{2^{i+1}}\delta^{b\mu_1\nu_1\mu_2\nu_2\ldots\mu_i\nu_i}_{a\alpha_1\beta_1\alpha_2\beta_2\ldots\alpha_i\beta_i}\prod^i_{r=1}R\indices{^{\alpha_r\beta_r}_{\mu_r\nu_r}}.
\end{equation}
By expanding the product in (2), it is easy to find the zeroth term is just a constant regarded as cosmological term, the first term is just Einstein-Hilbert term, and the second is the Gauss-Bonnet term as we mentioned before. But in fact because of the limitation of $ p $, we can find the Lovelock gravity equals to general relativity in four dimensions. In order to study the effect of higher order terms, we have to consider higher dimensional spacetime,i.g.five dimensions where the Lovelock gravity can contain the Gauss-Bonnet term but all other higher curvature terms in (1) vanish identically.\\
Inspired by the BHT new massive gravity in three dimensions \cite{Bergshoeff:2009hq}, the authors of Ref \cite{Oliva:2010eb} found by using some specific combinations of cubic curvature term in five dimensions, one can also get a theory like three-order Lovelock theory. Especially its equations of motion are of most second order in derivatives of metrics. The cubic term could be written as :
\begin{equation}
  I = \beta \int{\mathrm{d}^{5}x} \sqrt{-g}\mathcal{L} ,
\end{equation}
where $ \beta $ is a coupling constant and the Lagrangian is given by
\begin{equation}
 \mathcal{L}= \frac{7}{6} R\indices{^{ab}_{cd}}R\indices{^{ce}_{bf}}R\indices{^{df}_{ae}}
 + R\indices{_{ab}^{cd}}R\indices{_{cd}^{be}}R\indices{^{a}_{e}}
 +\frac12 R\indices{_{ab}^{cd}}R\indices{^{a}_{c}}R\indices{^{b}_{d}}
 -\frac13R\indices{^{a}_{b}}R\indices{^{b}_{c}}R\indices{^{c}_{a}}
 +\frac12 RR\indices{^{a}_{b}}R\indices{^{b}_{a}}
 +\frac{1}{12} R^3 .
\end{equation}
We can get the equations of motion by varying the action with respect to metric
\begin{equation} \begin{split}
  E_{ab} &=  \frac76[3R\indices{_{ahg}^{g}}R\indices{_{b}^{prd}}R\indices{_{pgr}^{h}}  -3\nabla_p\nabla_q(R\indices{^p_g^q_h}R\indices{_a^g_b^h}-R\indices{^p_{hbg}}R\indices{_a^{gqh}})-\frac12 g_{ab}R\indices{^{mn}_{cd}}R\indices{^{ce}_{nf}}R\indices{^{df}_{me}}] \\
   &+[R\indices{_{abcd}}R\indices{^{cspq}}R\indices{_{pqs}^{d}}-R\indices{_{a}^{qcd}}R\indices{_{cdb}^{h}}R\indices{_{qh}}
   +R\indices{_{b}^{dqc}}R\indices{_{adc}^{h}}R\indices{_{qh}}-\nabla_p\nabla_q(R\indices{_{ah}}R\indices{_{b}^{phq}}+R\indices{_{ah}}R\indices{_{b}^{hqp}}\\
 &\qquad+R\indices{_{bh}}R\indices{_{a}^{hqp}}+R\indices{_{bh}}R\indices{_{a}^{phq}}+R\indices{^q_h}R\indices{_a^h_b^p}+R\indices{^p_h}R\indices{_a^q_b^h}+
   \frac{1}{2}(g^{pq}R\indices{_a^{hcd}}R\indices{_{bhcd}}+g_{ab}R\indices{^{phcd}}R\indices{^q_{hcd}} \\
  &\qquad -g\indices{_a^p}R\indices{_b^{rcd}}R\indices{^q_{rcd}}-g\indices{_b^p}R\indices{_a^{rcd}}R\indices{^q_{rcd}}))-\frac12 g\indices{_{ab}}R\indices{_{mn}^{cd}}R\indices{_{cd}^{ne}}R\indices{^m_{e}}] \\
    &+\frac12[R\indices{_{ac}}R\indices{_b^{fcd}}R\indices{_{fd}}+2R\indices{_{acbd}}R\indices{_{cfdg}}R\indices{_{fg}}+\nabla_p \nabla_q(R\indices{_{ab}}R\indices{^{pq}}-R\indices{_a^p}R\indices{_b^q}+g^{pq}R\indices{_{abcd}}R\indices{^{cd}}\\
  &\qquad+g_{ab}R\indices{^{pcqd}}R\indices{_{cd}}-g\indices{_a^p}R\indices{^q_{cbd}}-g\indices{_B^p}R\indices{^q_{cad}}R\indices{^{cd}})-\frac12 g\indices{_{ab}}R\indices{_{mn}^{cd}}R\indices{^m_c}R\indices{^n_d}] \\
  &-\frac{1}{3}[3R\indices{_{acbd}}R\indices{^{ec}}R\indices{_e^d}+\frac23\nabla_p\nabla_q (g^{pq}R\indices{_a^c}R\indices{_{bc}}+g\indices{_{ab}}R\indices{^{ep}}R\indices{_e^q}-g\indices{_{b}^q}R\indices{^{qc}}R\indices{_{ac}}-g\indices{_a^p}R\indices{^{qc}}R\indices{_{bc}})-\frac12g\indices{_{ab}}R\indices{^m_n}R\indices{^n_c}R\indices{^c_m}] \\
  & +\frac12[R\indices{_{ab}}R\indices{^{cd}}R\indices{_{cd}}+2RR\indices{^{cd}}R\indices{_{acbd}}+\nabla_p \nabla_q(g\indices{_{ab}}g\indices{^{pq}}R\indices{^{cd}}R\indices{_{cd}}+g\indices{^{pq}}RR\indices{_{ab}}-g\indices{_a^q}g\indices{_b^q}R\indices{^{cd}}R\indices{_{cd}}+g\indices{_{ab}}RR\indices{^{pq}}\\
  &\qquad -g\indices{_b^p}RR\indices{_a^q}-g\indices{_a^p}RR\indices{_b^q})-\frac12RR\indices{^m_n}R\indices{^n_m}] \\
  &-\frac{1}{12}[3R^2 R\indices{_{ab}}+3\nabla_p \nabla_q (g\indices{_{ab}}g\indices{^{pq}}R^2-g\indices{_{a}^p}g\indices{^{q}_b}R^2)-g\indices{_{ab}}R^3].
\end{split} \end{equation}
Because of the appearance of two covariant derivatives, the rather complex field equations are fourth order in general. But we can find the trace of the field equation is proportional to its Lagrangian like the massive gravity as mentioned before. If we consider the static spherically symmetric metric
\begin{equation}
  ds^2 = -f(r)dt^2 + \frac{dr^2}{g(r)} + r^2d\Sigma^2_3 ,
\end{equation}
where $ d\Sigma_3 $ is the line element of an Euclidian three-dimensional space with constant curvature $ \gamma=\pm1,0 $, these gravitational field equations are then reduced to

\begin{equation}
\begin{split}
 E\indices{^r_r} &= -\frac{1}{2fr^6}(g-\gamma)^2(2fg-3rgf'-2\gamma f),\\
 E\indices{^t_t} &= -\frac{1}{2r^6}(g-\gamma)^2(2g-3g'r-2\gamma ), \\
 E\indices{^i_j} &= -\frac{\gamma -g}{4f^2r^6}[gr^2(\gamma - g)f'^2 + ff'r (5g'rg + 4\gamma g - 4g^2 -\gamma rg'),\\
  &\qquad + 4f^2 r (\gamma - g)g'- 2fgr^2 (\gamma -g)f''+ 4f^2(\gamma -g)^2] \delta\indices{^i_j}.
\end{split}
\end{equation}
where a prime denotes derivatives with respect to $r$, and indices $(i,j)$ run along the three dimensional spatial  manifold $ \Sigma_3 $. We can find these equations contain only second order derivatives of under these metric just like those in Lovelock gravity. By solving this system, one can find that an nontrivial solution of the theory is
\begin{equation}
  ds^2 = -(cr^{2/3}+ \gamma)dt^2 + \frac{dr^2}{cr^{2/3}+\gamma} +r^2d\Sigma^2_3.
\end{equation}
Further more it turns out that the Brikhoff's theorem remains valid in this theory \cite{Oliva:2010eb}.\\
In the remaining part of the paper, firstly we use the integration method \cite{Cai:2009qf} to show the cubic gravity also exist generalized Misner-Sharp energy. And then we pay our attention on the dynamical black hole solutions of this new cubic gravity. We follow the methods used in study of Lovelock gravity \cite{Cai:2008mh} to present a kind of similar generalized Vaidya spacetime solution in the cubic gravity. Finally we present the existence of the first law and the second law of black hole thermodynamics.

\section {Generalized Misner-Sharp energy in the cubic gravity}
 In order to compare with the Lovelock gravity, in this section we only consider the cubic term in five dimensions. We take the spherically symmetric space-time in the double-null form as \cite{Cai:2009qf}:
\begin{equation}
  ds^2=-2e^{-\varphi(u,v)}dudv+r^2(u,v)\gamma_{ij}dx^idx^j .
\end{equation}
The equations of gravitational field can be written as
\begin{equation}
  \beta E\indices{_{ab}}=8\pi G T\indices{_{ab}} ,
\end{equation}
where $ E\indices{_{ab}} $ is defined in (7). By using the metric in the double-bull coordinates, these equations can be written explicitly as
\begin{align}
  8\pi GT_{uu}&=-3\beta\frac{(1+2e^{\varphi}r_{,u}r_{,v})^2(r_{,u}\varphi _{,u}+r_{,uu})}{r^5},\\
  8\pi GT_{uv}&=-\beta e^{-\varphi}\frac{(1+2e^{\varphi}r_{,v}r_{,v})^2(1+2e^{\varphi}r_{,u}r_{,v}-3e^{\varphi}rr_{,uv})}{r^6}\\
  8\pi GT_{uu}&=-3\beta\frac{r_{,v}\varphi_{,v}+r_{,vv}(1+2e^{\varphi}r_{,v}r_{,u})^2}{r^5}.
\end{align}
If we can define a generalized Misner-Sharp energy, the gravitational field equations should be able to cast into the form(details could be found in the Ref \cite{Cai:2009qf} and we will review the process in the next section) :
\begin{equation}
  dE_{eff}=A\Psi_a dx^a +WdV ,
\end{equation}
where $ A=V_3r^3 $ and $ V=V_3r^4/4 $ are the area and volume of the 3-dimensional space with radius $r$, the $\Psi $ is the energy supply vector and $W$ is the energy density. We can express the right hand side as
\begin{equation}
  A\Psi_a dx^a +WdV=A(u,v)du+B(u,v)dv ,
\end{equation}
where
\begin{align}
A(u,v)&=V_3r^3e^{\varphi}(r_{,u}T_{uv}-r_{,v}T_{uu})\\
B(u,v)&=V_3r^3e^{\varphi}(r_{,v}T_{uv}-r_{,u}T_{vv}) .
\end{align}
The generalized Misner-Sharp energy can be derived by integrating the last equation if it is integrable. So the integrable condition has to be satisfied :
\begin{equation}
   \frac{\partial A(u,v)}{\partial v}=\frac{\partial B(u,v)}{\partial u} .
 \end{equation}
Just like the case in the Lovelock gravity, we can easily find that the integrable condition can be always satisfied by substituting the field equations in it.
Then the generalized Misner-Sharp energy expression could be found as :
\begin{align}
 E_{eff}&=\int A(u,v)du+\int [B(u,v)-\frac{\partial}{\partial v}\int A(u,v)du]dv \\
        &=\frac{\beta V_3(1+2e^{\varphi}r_{,u}r_{,v}) ^3}{16\pi Gr^2}   .
\end{align}
We can further express it in a covariant form as
\begin{align}
 E_{eff}&=\frac{\beta V_3(1-h\indices{^{ab}}D_arD_br) ^3}{16\pi Gr^2} .
 \end{align}
As done in the the Gauss-Bonnet gravity\cite{Cai:2009qf}, in this section we get the generalized Misner-Sharp energy for the new cubic gravity term. If we compare the field equations and Misner-Sharp energy with the contribution of the cubic term in the Lovelock gravity, we can find they are very familiar in form regardless of the coefficients. But in fact as we all know in five dimensions the cubic term in Lovelock theory can't give any contribution to the field equations where the coefficients of the cubic term always vanish. This is why we consider the new cubic gravity in five dimensions which is interesting due to Ads/CFT correspondence, we can regard it as the extension of  Lovelock theory in five dimensions (of course in the complete theory we should add the cosmological term, Hilbert-Einstein term and Gauss-bonnet term together).\\

\section {Generalized Vaidya spacetime in the new cubic gravity}
\subsection {Generalized Vaidya solution }
Vaidya found an exact solution of the Einstein equation field equations with radiation matter in four dimensions:
\begin{equation}
  ds^2=-[1-\frac{2M(v)}{r}]dv^2+2dvdr+r^2d\theta^2 +r^2{\sin \theta}^2d\phi^2,
\end{equation}
where $M(v)$ is an arbitrary function of $v$. It can be regarded as a non-stationary Schwarzschild spacetime where there is an outgoing spherically symmetric radiation of massless particles. The apparent horizon is located at $ r=2M(v)$, the energy momentum tensor is given by $T_{ab}=\mu l_al_b$, where $ l_a=(1,0,0,0) $ and $\mu$ is the energy density of the radiation matter.
In this section we want to show there is also a kind of generalized solutions in the new cubic gravity and  study the thermodynamics of the apparent horizon of this dynamical black hole in five dimensions. First we assume a five-dimensional metric similar as (24)
\begin{equation}
  ds^2=-f(v,r)dv^2+2dvdr+r^2d\Omega^2_3 ,
\end{equation}
and we can express the components of $H\indices{_{ab}}$ in gravitational field equations (9) as:
\begin{align}
H\indices{^v_v} &= \frac{(-1+f)^2(2-2f+3rf_{,r})}{2r^6},\\
H\indices{^r_r} &=H\indices{^v_v},\\
H\indices{^r_v} &=-\frac{3(-1+f)^2f_{,v}}{2r^5}, \\
H\indices{^i_j} &=\frac{1}{r^6}(-1+f)\big(2+2f^2+4rf_{,r}+2r^2f_{,r}^2-r^2f_{,rr}+f(-4-4rf_{,r}+r^2f_{,rr})\big)\delta\indices{^i_j}.
\end{align}
From these field equations it is explicit that they are only of most second order in derivatives which is a motivation for us to consider the new cubic gravity. In fact if we define a new function $F(v,r)=\frac{1-f(v,r)}{r^2}$ then we can simplify these equations as:
\begin{align}
H\indices{^v_v} &=H\indices{^r_r} =  -\frac{1}{2r^3}\frac{\partial (F^3r^4)}{\partial r}, \\
H\indices{^r_v} &=\frac{r}{2}\frac{\partial F^3}{\partial v},   \\
H\indices{^i_j} &= -\frac{\delta^i_j}{2r^3}\frac{\partial^2 (F^3r^4)}{\partial^2 r} .
\end{align}
We don't really regard the third equation as an independent one, because it's easy to find the relation
\begin{equation}
H\indices{^i_j}=\delta^i_j(r\frac{\partial H\indices{^v_v}}{\partial r}\frac13 +H\indices{^v_v}).
\end{equation}
In order to find some solutions of the field equations, we should priorly assume a reasonable energy tensor. Of course the simplest assumption is the same case as in the Vaidya spacetime.\\
$\bullet$ case 1 \\
In this case, we take the energy momentum tensor as $T_{ab}=\mu l_al_b$, where $l_a=(1,0,0,0)$.
So we have
\begin{equation}
  \beta H\indices{^r_v} =8 \pi G \mu \qquad \qquad H\indices{^v_v}= H\indices{^r_r}=0=H\indices{^i_j}
\end{equation}
Solving these equations, we can obtain
\begin{align}
\beta F^3 &= \frac{16 \pi Gm(v)}{\Omega_3 r^4},  \\
 8\pi G \mu &= -\beta \frac{r}{2}\frac {\partial F^3}{\partial v}= \frac{8\pi G}{r^3}\frac{\partial m(v)}{\partial v},
\end{align}
where $m(v)$ is an arbitrary positive function of $v$, which appears as an integration constant. Eq(36) tells us with
\begin{equation}
\mu =\frac{m_{,v}}{\Omega_3r^3}.
\end{equation}
\\
$\bullet$ case 2 \\
We can generalize the last solution to a more general case by assuming the energy momentum tensor has the form $T\indices{^i_j}=\sigma T\indices{^v_v}=\sigma T\indices{^r_r}$. In addition to the field equations, we also have the constraint $\nabla_a T\indices{^a_b}=0$ which can be explicitly expressed as
\begin{align}
b=v,\qquad \partial_vT\indices{^v_v}+\partial_rT\indices{^r_v}+\frac{3}{r}T\indices{^r_v} &=0, \\
b=r, \qquad \partial_rT\indices{^r_r}+\frac{3}{r}(1-\sigma )T\indices{^r_r} &= 0.
\end{align}
We can get
\begin{align}
T\indices{^r_r}=T\indices{^v_v}=-C(v)r^{3(\sigma -1)},
\end{align}
where $C(v)$ is a function of $v$. So from the field equations, we have
\begin{equation}
8\pi GC(v)r^{3\sigma} =\frac{\beta}{2}\frac{\partial (F^3r^4)}{\partial r}.
\end{equation}
with the solution
\begin{equation}
  \beta F^3=16\pi G[\frac{m(v)}{\Omega_3 r^4}+\frac{C(v)\Theta (r)}{r^4}].
\end{equation}
The function $\Theta(r)$ is defined as
\begin{equation}
\Theta(r)=\int dr r^{3\sigma}= \begin{cases}
\ln r , & \sigma =-1/3 \\
\frac{r^{3\sigma+1}}{3\sigma +1}, & \sigma \neq-1/3
\end{cases}
\end{equation}
Considering the field equation $8\pi G \widetilde{\mu} = -\beta \frac{r}{2}\frac {\partial F^3}{\partial v}$, we can find
\begin{equation}
  T\indices{^r_v}=\widetilde{\mu}=\frac{\partial_v m(v)}{\Omega_3r^3}+\frac{\partial_vC(v)\Theta(r)}{r^3}.
\end{equation}
The first term is just what we got in the case 1. The generalized Vaidya solution is very similar to the contribution of the cubic term in the Lovelock gravity except for the difference in the factor.
In order to be beneficial for us to study the thermodynamics of the black hole in next section, we express the energy momentum tensor in a covariant form as
\begin{equation}
  T_{ab}=\widetilde{\mu}l_al_b-P(l_an_b+n_al_b)+\sigma Pq_{ab},
\end{equation}
where $ P= T^r_r =-C(v)r^{3(\sigma -1)} $, the $n_a$ is the null vector which satisfies $l_an^a=-1$, $q_{ab} $ is the projection operator defined as $q_{ab}=g_{ab}+l_an_b+l_bn_a$. Thus the metric can be written as $g_{ab}=h_{ab}+q_{ab}$, where the metric $h_{ab}$ of 2-dimensional spacetime transverse to the 3-dimensional sphere is $-(l_an_b+l_bn_a)$. In the coordinates (25), we have
\begin{equation}
  l_a=(1,0,0,0,0) ,\qquad and \qquad  n_a=(f/2,-1,0,0,0) .
\end{equation}
\subsection {Thermodynamics on the apparent horizon}
In this section we will focus on the apparent horizon of the black hole in the dynamical Vaidya spacetime rather than the event horizon of stationary black hole, which is the boundary of the past of future infinity. For a dynamical black hole, the dynamical horizon and the outer trapping horizon are not null hypersurfaces but spacelike hypersurfaces. So the Wald entropy formula \cite{Wald:1993nt} associated with event horizon may not be applicable. Before we discuss the entropy and the energy associated with apparent horizon, we firstly review some works of Hayward \cite{Hayward:1993wb,Hayward:1997jp,Hayward:1998ee} which is focused on four-dimensional Einstein gravity. From these works, we can find the deep relation between the equations of motion of gravitational fields and thermodynamics of spacetimes.\\
Consider an $n$-dimensional spherically symmetric spacetime whose metric is in the double null form
\begin{equation}
  ds^2=h_{ab}dx^adx^b +r^2(x)d\Omega^2_{n-2} ,
\end{equation}
where ${x^a}$ are coordinates of the two-dimensional spacetime $(M,h_{ab})$ which is transverse to the $(n-2)$-dimensional sphere.
Then we need to define three important physical quantities : the work density $W=-1/2h^{ab}T_{ab}$ which corresponds to the work term in the first law, and the energy supply $\Psi_a=T^b_a\partial_b r +W\partial_a r$, and the Misner-Sharp energy inside the sphere with radius $(r)$ :
\begin{equation}
  E=\frac{(n-2)}{16\pi}\Omega_{n-2}r^{n-3}(1-h^{ab}\partial_ar\partial_br) .
\end{equation}
By using these quantities, we can put some components of the Einstein equations into the form called unified first law:
\begin{equation}
  dE=A\Psi+WdV ,
\end{equation}
where $A=\Omega_{n-2}r^{n-2}$ and $V=\Omega_{n-2}r^{n-1}/(n-1)$.
Defining a new operator $\delta $ as: take a Lie derivatives with respect to $\xi$ and then evaluate it on the apparent horizon.
The vector $\xi$ is tangent to the trapping horizon of the spacetime. And Hayward showed that on the trapping horizon, one can find the relation:
\begin{equation}
  A\Psi_adx^a=\frac{\kappa}{8\pi}\mathcal{L}_{\xi}A=\frac{\kappa}{2\pi}\mathcal{L}_{\xi}S=T\delta S ,
\end{equation}
where $S=A/4$, $T=\kappa/2\pi$, surface gravity $\kappa=D^aD_ar/2$, the covariant derivative operator $D$ is associated with metric $h_{ab}$.
The last equation can be understood as the Clausius relation $\delta Q=T\delta S$ with $\delta Q=  A\Psi_adx^a$.
After projecting the unified first law onto the trapping horizon, we can get the first law of thermodynamics :
\begin{equation}
  \delta E=T\delta S +W\delta V
\end{equation}
But these results are only applicable for Einstein gravity i.e., the case without any higher order curvature corrections in this action. Following the Refs \cite{Cai:2006rs,Cai:2008mh} in order to deal with these corrections terms like the Lovelock terms or others, we can rewrite the complete equations of motion $\mathcal{G}_{ab}=8\pi T_{ab}^{(m)}$ into the standard form :
 \begin{equation}
   G_{ab}=8\pi T_{ab}=8\pi T_{ab}^{(m)}+8\pi T_{ab}^{(e)},
 \end{equation}
where $T_{ab}^{(e)}=-\mathcal{G}_{ab}+G_{ab}$ can be regarded as effective energy momentum tensor coming from the corrections term in the action.
Then in the two-dimensional spacetime $(M^2,h_{ab})$ we can define the similar work term and the energy supply :
\begin{align}
    W &= -1/2h^{ab}T_{ab}=W^{(m)}+W^{(e)}, \\
\Psi_a&= T^b_a\partial_b r +W\partial_a r=\Psi_a^{(m)}+\Psi_a^{(e)}.
\end{align}
From these definitions, we can rewrite the unified first law as
\begin{equation}
 dE-A\Psi^{(e)}-W^{(e)}dV=A\Psi^{(m)}+W^{(m)}dV .
\end{equation}
The left hand side of the equation is totally determined by geometry and the right hand side is about the matter . If one also wants to get a generalized fist law and the first law of thermodynamics for a new gravity theory, one should deal with two things : \\
Firstly, the left hand side should could be written into a totally derivative form $ dE_L $, then by this way we can get the generalized Misner-Sharp energy $E_L$. In the last section, we directly use this way to get the Misner-Sharp energy (23).\\
Secondly, on apparent horizon $ A\Psi^{(m)}_a\xi^a=\kappa/(8\pi)\delta A-A\Psi^{(e)}_a\xi^a$ should be able to cast into a form $T\delta S $ of the Clausias relation just as in Einstein gravity.\\
Of course not all gravity theories can be able to realize the two points, for example the $f(R)$ theory and scalar -tensor theory are two examples\cite{Cai:2006rs}. But in the following content we can find that the new cubic theory can be done in this way just like the Lovelock theory.\\
In the generalized Vaidya spacetime (25) where the horizon is located at $r_A$ determined by $f(v,r_A)=0$, it is easy to find $\kappa =\partial _r f(v,r)$,
\begin{align}
    W = -T^{r(m)}_r&=-P=C(v)r^{3(\sigma-1)} \\
\Psi_a^{(m)}&= \widetilde{\mu}l_a= -T^{r(m)}_vl_a ,
\end{align}
and check that these quantities satisfy the unified first law. On the apparent horizon,
we should set a tangent vector $\xi_a$ on the horizon as
\begin{equation}
-\xi_a= (\frac{\partial_vf}{2\kappa})l_a+n_a ,
\end{equation}
which is opposite to the case in \cite{Cai:2008mh}.
The first one can be easily checked by substituting the field equations for the cubic theory to the equation (55), hence the function $E_L$ can be expressed as
\begin{align}
E_L &=\beta\frac{\Omega_3}{16\pi}\frac{(1-f)^3}{r^2} \\
    &=\beta\frac{\Omega_3}{16\pi}\frac{(1-h^{ab}D_arD_br)^3}{r^2}.
\end{align}
The covariant form is explicitly same with what we got in the previous section.
This means that we can also get a generalized unified first law in the Vaidya spacetime for the  new cubic theory:
\begin{equation}
  dE_L=A\Psi^{(m)}+W^{(m)}dV .
\end{equation}
In order to study the thermodynamics on the apparent horizon, we start with the Clausius-like equation:
\begin{equation}
A\Psi_a\xi^a=A\Psi^{(m)}_a\xi^a+A\Psi^{(e)}_a\xi^a=\frac{\kappa}{8\pi}\delta A.
\end{equation}
We should emphasize that the heat flow $\delta Q$ should be determined by the matter energy momentum tensor, so it should be defined as:
\begin{equation}
  \delta Q \equiv A\Psi^{(m)}_a\xi^a=A\widetilde{\mu}l_a\xi^a= A\widetilde{\mu}
\end{equation}
According to the result we got in the previous section, it is easy to find
\begin{equation}
  -2A\widetilde{\mu}=\beta\frac{3A\partial_vf}{8\pi r^5},
\end{equation}
\begin{equation}
  \delta Q=A\widetilde{\mu}=-\frac{\kappa}{2\pi} (\frac{\partial_vf}{2\kappa})\frac{3A\beta }{4r^5}.
\end{equation}
If we define the temperature by $T=\kappa/2\pi$ and the entropy of the apparent horizon by
\begin{equation}
  S=-\beta\frac{3A}{4r^4},
\end{equation}
then we can get the Calusius relation $\delta Q= T\delta S $ which is required by the first condition.
After projecting the generalized first law (61) on the apparent horizon with $W^{(m)}=-P$, we obtain first law of the apparent horizon:
\begin{equation}
  \delta E_L=T\delta S -P\delta V ,
\end{equation}
which can be understood as a version of physical process for the first law of black hole thermodynamics i.e., an active version of the first law.\\
On the other hand we can understand the first law from a passive version by considering a small perturbation to state of the dynamical black hole. We use the operator $" \Delta "$ to denote the variation of two nearby points in the solution space.e.g, $r_A$ and $r_A+\Delta r_A$. With considering the condition $f(r_A,v)=0$ on the horizon, the temperature $\kappa$, energy inside the apparent horizon $M_L$ and the pressure $P$ on it are, respectively,
\begin{align}
\kappa &=\frac{1}{6\beta}(-2\beta \frac{1}{r_A}+16\pi C(v)r_A^{(3\sigma +2)}),\\
M_L    &=\beta\frac{\Omega_3}{16\pi}r_A^{-2},                             \\
P      &=-C(v)r_A^{3(\sigma-1)},                                      \\
 S     &=-\beta\frac{3A}{4r_A^4}.
\end{align}
Then it is easy to show the first law in passive version
\begin{equation}
\Delta M_L=T\Delta S-P\Delta V
\end{equation}
is satisfied.
\section{Some Discussions}
\subsection {The complete cubic gravity theory in five dimensions}
As we said at the beginning of the article, the motivation for us to consider such kind of cubic term is to extend the Lovelock gravity in five dimensions. So the cubic gravity should have many similarities with the Lovelock gravity. Firstly, we can find their gravitational field equations are all at most second order in derivatives in spite of the complexity of their actions. Secondly, we can also find the similarities in the thermodynamics of solutions in these two theories. So in order to make it clear, we can compare these results with those in the Lovelock theory in arbitrary spacetime dimensions $n$ \cite{Cai:2008mh}.\\
The Lovelock theory exists the generalized Vaidya spacetime solutions\cite{Cai:2008mh}:
\begin{gather}
  \sum_i^pc_i\frac{(n-2)!}{(n-2i-1)!}F^i=16\pi G\Bigl( \frac{m(v)}{\Omega_{n-2}r^{n-1}}+\frac{C(v)\Theta(r)}{r^{n-1}} \Bigr),  \\
  T\indices{^r_v}=\widetilde{\mu}=\frac{\partial_vm(v)}{\Omega_{n-2}r^{n-2}}+\frac{\partial_vC(v)\Theta(v)}{r^{n-2}},
\end{gather}
where $p=[(n-1)/2]$, $c_i$ are arbitrary constants with dimension of $[length]^{2i-1}$  \\
Associated Thermodynamics functions are:
\begin{align}
\text{entropy of apparent horizon} &:\qquad    S=\frac{A}{4}\sum_i^pc_i\frac{i(n-2)!}{(n-2i)!}r^{2-2i}_A ,\\
\text{generalized Misner-sharp energy} &: \qquad E_L=\frac{\Omega_{n-2}}{16\pi}\sum_i^p\frac{c_i(n-2)!}{(n-2i-1)!}r^{n-2i-1}(1-f(v,r))^i,\\
\text{the first law of apparent horizon } &:\qquad \delta E_L=T\delta S -P\delta V.
\end{align}
From these expressions of the cubic term,i.e. $i=3$ term, one can find they are almost same with what we got in the previous section for the cubic gravity. So in fact after a replacement as :
\begin{equation}
  c_i\frac{(n-2)!}{(n-2i-1)!}\mid_{i=3} \Longrightarrow \beta,
\end{equation}
and taking spacetime dimension as 5, these results are actually same. But we should stress that it doesn't mean the two theories are equivalent because the factor $c_i\frac{(n-2)!}{(n-2i-1)!}$ is not meaningful for $n=5,i=3$  because of the limitations $p=[(n-1)/2]$, for Lovelock theory. So we can consider the new cubic theory (2) as a good extension for Lovelock theory in five dimensional spherically spacetime.\\
One natural question is why these two theories are so similar? Maybe the answer has been given in the construction of the cubic term by the authors of \cite{Oliva:2010eb} who consider the new theory initially in order to present another linearly independent cubic curvature invariant that has same property with pure quadratic BHT in three dimensions. The cubic curvature invariant (2) can be written in two parts:
\begin{align}
\begin{split}
\mathcal{L}:\frac{1}{2^k}(\frac{1}{D-2k+1})\delta^{a_1b_1\cdots a_kb_k}_{c_1d_1 \cdots  c_kd_k}(C\indices{^{c_1d_1}_{a_1b_1}}\cdots  C\indices{^{c_kd_k}_{a_kb_k}}-R\indices{^{c_1d_1}_{a_1b_1}}\cdots R\indices{^{c_kd_k}_{a_kb_k}}) \\
-\alpha_kC\indices{^{a_kb_k}_{a_1b_1}}C\indices{^{a_1b_1}_{c_2b_2}}\cdots C\indices{^{a_{k-1}b_{k-1}}_{a_kb_k}},
\end{split}
\end{align}
where the $C_{abcd}$ is the Wely tensor and
\begin{equation*}
  \alpha_k=\frac{(D-4)!}{(D-2k+1)!}\frac{[k(k-2)D(D-3)+k(k+1)(D-3)+(D-2k)(D-2k-1)]}{[(D-3)^{k-1}(D-2)^{k-1}+2^{k-1}-2(3-D)^{k-1}]}.
\end{equation*}
The first part invariant containing the $k$th Lovelock invariant and a term proportional to $(Wely)^k$, can give field equations whose trace is of second order in all dimensions \cite{Oliva:2010zd}. And the second part is an appropriate multiple of $(Wely)^k$. Then substraction of the two terms can make all the components of the field equations arising from the invariant (79) to be of second order. If we set $D=5,k=3$ then we can get the invariant (6). So we can find the reason for similarity between the new cubic term and Lovelock
just is the construction of (79) which only retains the contributions of the cubic Lovelock term and the numerical factor $1/(D-2k-1)$ which makes the final result meaningful rather than to be zero contributing by  the cubic Lovelock term in five dimensions .\\
In the previous sections, we have only considered the cubic term (2) for simplicity. Of course we can supply the $i=0,1,2$ Lovelock term, i.e., the cosmological term, Hilbert-Einstein term and the Gauss-Bonnet term and obtain a complete theory in five dimensions as the extension of the Lovelock theory.\\
As a result, we have
\begin{equation}
  S=\frac{1}{16\pi G} \int d^5x  \sqrt{-g}\big((R-2\Lambda+\alpha L_{GB}+\beta L{cu})+L_m \big),
\end{equation}
where $L_{GB}=R^2-4R_{\mu\nu}R^{\mu\nu}+R_{\mu\nu\rho\sigma}R^{\mu\nu\rho\sigma}$ is the Gauss-Bonnet term, $L_{cu}$ is the new cubic term (2), $L_m$ is the lagrangian of matter. By varying the action with respect to metric, we can get the gravitational field equations:
\begin{equation}
  8\pi GT_{\mu\nu}=G_{\mu\nu} +\alpha M_{\mu\nu} +\Lambda g_{\mu\nu} +\beta H_{\mu\nu} ,
\end{equation}
where the $H_{\mu\nu}$ is given by (7) and
\begin{align}
 G_{\mu\nu}&=R_{\mu\nu}-\frac{1}{2}Rg_{\mu\nu}, \\
 M_{\mu\nu}&=2(RR_{\mu\nu}-2R_{\mu\alpha}R_{\nu}^{\alpha}-2R^{\alpha\beta}R_{\mu\alpha\nu\beta}+R\indices{_{\mu}^{\alpha\beta\gamma}}R_{\nu\alpha\beta\gamma})
 -\frac12g_{\mu\nu}L_{GB} .
\end{align}
And  we can get the following results by parallel analysis about thermodynamics between the Lovelock theory and new cubic term:
\begin{align}
\text{entropy of apparent horizon} &:   S=\frac{A}{4}(1+12\alpha r^{-2}_A -3\beta r^{-4}_A ) \\
\text{generalized Misner-sharp energy} &: E_L=\frac{\Omega_{3}}{16\pi}[-\frac{1}{2}\Lambda r^4+3r^2(1-f)+6\alpha (1-f)^2+\beta r^{-2}(1-f)^3]\\
\text{the first law of apparent horizon } &:\delta E_L=T\delta S -P\delta V
\end{align}\\

\subsection {The second law.}
Recently, Wall show presents that the second law of black hole thermodynamics holds for arbitrarily complicated theories of higher curvatures gravity when considering only linearized perturbations to stationary black hole \cite{wall2015second}.
In this subsection we would like to  to check if the cubic theory obeys the second law of thermodynamics in the generalized Vaidya spacetime. We should stress the Vaidya spacetime is not stationary, so it is not contained in Wall's conclusion. But this kind of spacetime would make us able to directly use the positive version to consider the variation of black hole entropy rather than perturbation. We can find not only the Lovelock theory but also the new cubic theory obeys the second law. The key point is that the null energy condition(NEC) behaves as in the Einstein-Hilbert theory where the NEC implies that the area of any future event horizon is always increasing.(Wall has found the $m$th term in Lovelock gravity in $2m$ dimensions can violate the classical second law because of the abrupt change of entropy within the merger of two black hole\cite{Sarkar:2010xp}. But this special condition doesn't happen in our case.)\\
We start from the Lovelock gravity. From the entropy of apparent horizon (75), we can get
\begin{equation}
\frac{\partial S}{\partial v}=\frac{\Omega_{n-2}}{4}\sum_i^pc_i\frac{i(n-2)!}{(n-2i-1)!}r^{n-2i-1}_A \frac{\partial r_A}{\partial v} .
\end{equation}
Then we consider the NEC.i.e., $T_{ab}n^an^b \geq 0$, where the $n^a$ is arbitrary null vector. By using the covariant form (45), it is easy to show that
\begin{equation}
  T_{ab}n^an^b = \widetilde{\mu} \geq 0 .
\end{equation}
From  equations of gravitational field, we can have
\begin{align}
\sum_i^pc_i\frac{(n-2)!}{2(n-2i-1)!}F^i=16\pi G \bigg(\frac{m(v)}{\Omega_{n-2}r^{n-1}}+\frac{C(v)\Theta(r)}{r^{n-1}}\bigg)
\end{align}
\begin{align}
8\pi G \widetilde{\mu}&= \sum_i^pc_i\frac{i(n-2)!}{2(n-2i-1)!} \frac{\partial F^i}{\partial v} \\
T^r_v=\widetilde{\mu}&=\frac{\partial_v m(v)}{\Omega_{n-2}r^{n-2}}+\frac{\partial_v C(v)\Theta (r)}{r^{n-2}} \geq 0
\end{align}
Then considering the field equation (73) on the horizon $r_A$, we have
\begin{equation}
  \sum_i^p c_i \frac{(n-2)!}{(n-2i-1)!}r_A^{n-2i-1} = 16 \pi G \bigg( \frac{m(v)}{\Omega_{n-2}}+C(v)\Theta(r_A) \bigg) .
\end{equation}
After taking differentiation  with respect to the $v$, we can get
\begin{equation}
   \sum_i^p c_i \frac{(n-2)!}{(n-2i-2)!}r_A^{n-2i-2} \frac{\partial r_A}{\partial v} = 16 \pi G \bigg( r_A^{n-2}\widetilde{\mu} +C(v)r_A^{(n-2)\sigma} \frac{\partial r_A}{\partial v} \bigg) ,
\end{equation}
and further
\begin{equation}
   (\sum_i^p c_i \frac{(n-2)!}{(n-2i-2)!}r_A^{n-2i-2}- 16 \pi GC(v)r_A^{(n-2)\sigma} ) \frac{\partial r_A}{\partial v} = 16 \pi G  r_A^{n-2}\widetilde{\mu} \geq  0 .
\end{equation}
Then considering the  component $ T^v_v $  of equations of gravitational field:
\begin{equation}
  \sum_i^p c_i \frac{(n-2)!}{2(n-2i-1)!} \frac{\partial (r^{n-1}F^i)}{\partial r} = 8\pi G C(v)r^{(n-2)\sigma },
\end{equation}
we have a useful form on the horizon:
\begin{equation}
  \sum_i^p c_i \frac{(n-2)!}{2(n-2i-2)!}r_A^{n-2i-2} +\sum_i^p c_i \frac{i(n-2)!}{2(n-2i-1)!}(-\frac{\partial f}{\partial r})\big |_{r=r_A}r_A^{n-2i-1} = 8\pi G C(v)r_A^{(n-2)\sigma }.
\end{equation}
(If $ n-2i-2 < 0$, this term is just 0)\\
Substituting (94) to (96),we obtain:
\begin{equation}
  \sum_i^p c_i \frac{i(n-2)!}{(n-2i-1)!}(+\frac{\partial f}{\partial r}) \bigg |_{r=r_A}r_A^{n-2i-1}\frac{\partial r_A}{\partial v}= 16 \pi G  r_A^{n-2}\widetilde{\mu} \geq  0 .
\end{equation}
The left hand side of last equation is actually $ T \partial_v S$, so we have proved the second law:
\begin{equation}
  \frac{\partial S}{\partial v} \geq 0.
\end{equation}
%We only consider the temperature $T=\kappa /2\pi > 0$. If the surface gravity is negative, the temperature can be defined as $T'=-\kappa /2\pi $ , then the entropy $S'$ would be $ -S $, which doesn't change our conclusion (98).%
Further we can use the same process to prove the second law of thermodynamics for the single cubic gravity and complete cubic gravity in five dimensions. What we need to do is to take spacetime dimension $n$ as 5 and to do the replacement (78).  As we mention in the beginning of the subsection, the key point of the proof is the NEC.\\
In the end, we want to stress our motivation is to consider the cubic gravity as the extension of Lovelock gravity in five dimensions and as an useful concrete model for the Ads/CFT correspondence in 5 dimensions. At the same time with the paper \cite{Oliva:2010eb}, the authors of \cite{Myers:2010ru} also independently found a equivalent cubic invariant by using another method and studied the black solutions in the theory. They also gave the physical constraints on the couplings in the gravitational theory, investigated hydrodynamics aspects of the dual gauge theory, and also examined holographic version of the $c$ theorem in \cite{Myers:2010jv,Myers:2010tj}.

\section*{Acknowledgements}
We thank Rong-Gen Cai for useful guidance and discussion.

\bibliography{mybiography}   % 说明使用的文献数据库是 mybiography.bib

\begin{thebibliography}{10}

\bibitem{Zumino:1985dp}
Bruno Zumino.
\newblock {Gravity Theories in More Than Four-Dimensions}.
\newblock {\em Phys. Rept.}, 137:109, 1986.

\bibitem{Gross:1986iv}
David~J. Gross and Edward Witten.
\newblock {Superstring Modifications of Einstein's Equations}.
\newblock {\em Nucl. Phys.}, B277:1, 1986.

\bibitem{Garraffo:2008hu}
Cecilia Garraffo and Gaston Giribet.
\newblock {The Lovelock Black Holes}.
\newblock {\em Mod. Phys. Lett.}, A23:1801--1818, 2008.

\bibitem{Antoniadis:1997eg}
Ignatios Antoniadis, S.~Ferrara, R.~Minasian, and K.~S. Narain.
\newblock {$R^4$ couplings in M and type II theories on Calabi-Yau spaces}.
\newblock {\em Nucl. Phys.}, B507:571--588, 1997.

\bibitem{Lovelock:1971yv}
D.~Lovelock.
\newblock {The Einstein tensor and its generalizations}.
\newblock {\em J. Math. Phys.}, 12:498--501, 1971.

\bibitem{Bergshoeff:2009hq}
Eric~A. Bergshoeff, Olaf Hohm, and Paul~K. Townsend.
\newblock {Massive Gravity in Three Dimensions}.
\newblock {\em Phys. Rev. Lett.}, 102:201301, 2009.

\bibitem{Oliva:2010eb}
Julio Oliva and Sourya Ray.
\newblock {A new cubic theory of gravity in five dimensions: Black hole,
  Birkhoff's theorem and C-function}.
\newblock {\em Class. Quant. Grav.}, 27:225002, 2010.

\bibitem{Cai:2009qf}
Rong-Gen Cai, Li-Ming Cao, Ya-Peng Hu, and Nobuyoshi Ohta.
\newblock {Generalized Misner-Sharp Energy in f(R) Gravity}.
\newblock {\em Phys. Rev.}, D80:104016, 2009.

\bibitem{Cai:2008mh}
Rong-Gen Cai, Li-Ming Cao, Ya-Peng Hu, and Sang~Pyo Kim.
\newblock {Generalized Vaidya Spacetime in Lovelock Gravity and Thermodynamics
  on Apparent Horizon}.
\newblock {\em Phys. Rev.}, D78:124012, 2008.

\bibitem{Wald:1993nt}
Robert~M. Wald.
\newblock {Black hole entropy is the Noether charge}.
\newblock {\em Phys. Rev.}, D48:3427--3431, 1993.

\bibitem{Hayward:1993wb}
S.~A. Hayward.
\newblock {General laws of black hole dynamics}.
\newblock {\em Phys. Rev.}, D49:6467--6474, 1994.

\bibitem{Hayward:1997jp}
Sean~A. Hayward.
\newblock {Unified first law of black hole dynamics and relativistic
  thermodynamics}.
\newblock {\em Class. Quant. Grav.}, 15:3147--3162, 1998.

\bibitem{Hayward:1998ee}
Sean~A. Hayward, Shinji Mukohyama, and M.~C. Ashworth.
\newblock {Dynamic black hole entropy}.
\newblock {\em Phys. Lett.}, A256:347--350, 1999.

\bibitem{Cai:2006rs}
Rong-Gen Cai and Li-Ming Cao.
\newblock {Unified first law and thermodynamics of apparent horizon in FRW
  universe}.
\newblock {\em Phys. Rev.}, D75:064008, 2007.

\bibitem{Oliva:2010zd}
Julio Oliva and Sourya Ray.
\newblock {Classification of Six Derivative Lagrangians of Gravity and Static
  Spherically Symmetric Solutions}.
\newblock {\em Phys. Rev.}, D82:124030, 2010.

\bibitem{wall2015second}
Aron~C Wall.
\newblock A second law for higher curvature gravity.
\newblock {\em arXiv preprint arXiv:1504.08040}, 2015.

\bibitem{Sarkar:2010xp}
Sudipta Sarkar and Aron~C. Wall.
\newblock {Second Law Violations in Lovelock Gravity for Black Hole Mergers}.
\newblock {\em Phys. Rev.}, D83:124048, 2011.

\bibitem{Myers:2010ru}
Robert~C. Myers and Brandon Robinson.
\newblock {Black Holes in Quasi-topological Gravity}.
\newblock {\em JHEP}, 08:067, 2010.

\bibitem{Myers:2010jv}
Robert~C. Myers, Miguel~F. Paulos, and Aninda Sinha.
\newblock {Holographic studies of quasi-topological gravity}.
\newblock {\em JHEP}, 08:035, 2010.

\bibitem{Myers:2010tj}
Robert~C. Myers and Aninda Sinha.
\newblock {Holographic c-theorems in arbitrary dimensions}.
\newblock {\em JHEP}, 01:125, 2011.

\end{thebibliography}

\end{document}